\documentclass[12pt,prb,aps,preprint]{revtex4}

\usepackage{graphics}
\usepackage{graphicx}
\usepackage{amsfonts}
\usepackage{amsmath}
\usepackage{float}
\usepackage{epsfig}

\begin{document}
\title{Effect of Dimensionality on the Continuum Percolation of Overlapping Hyperspheres and Hypercubes:
II. Simulation Results and Analyses}

\author{S. Torquato}

\email{torquato@electron.princeton.edu}

\affiliation{\emph{Department of Chemistry, Department of Physics, Princeton Center for Theoretical Science,
Princeton Institute for the Science and Technology of
Materials, and Program in Applied and Computational Mathematics}, \emph{Princeton University},
Princeton NJ 08544}

\author{Y. Jiao}

\email{yjiao@princeton.edu}

\affiliation{\emph{Princeton Institute for the Science and Technology of
Materials}, \emph{Princeton University},
Princeton NJ 08544}

\begin{abstract}
In the first paper of this series [S. Torquato, J. Chem. Phys. {\bf 136}, 054106 (2012)],
analytical results concerning the continuum percolation of overlapping hyperparticles
in $d$-dimensional Euclidean space $\mathbb{R}^d$ were obtained, including lower bounds
on the percolation threshold. In the present investigation, we provide additional
analytical results for certain cluster statistics, such as  the concentration of $k$-mers and related
quantities, and obtain an upper bound on the percolation threshold $\eta_c$. We utilize the tightest lower bound obtained in the first paper
to formulate an efficient simulation method, called  the {\it rescaled-particle} algorithm, to estimate continuum percolation properties
across many space dimensions with heretofore unattained accuracy.  This simulation procedure is applied to compute the threshold $\eta_c$
and associated mean number of  overlaps per particle ${\cal N}_c$ for
both overlapping hyperspheres and oriented hypercubes for $ 3 \le d \le 11$.
These simulations results are compared to corresponding upper and lower bounds on these
percolation properties. We find that the bounds converge to one another
as the space dimension increases, but the lower bound provides an excellent
estimate of $\eta_c$ and ${\cal N}_c$, even for relatively low dimensions. We confirm a prediction
of the first paper in this series that low-dimensional percolation
properties encode high-dimensional information. We also show that the concentration
of monomers dominate over concentration values  for higher-order clusters (dimers, trimers, etc.) as
the space dimension becomes large. Finally, we provide accurate analytical
estimates of the pair connectedness function and blocking function
at their contact values for any $d$  as a function of density.

\end{abstract}


\maketitle
\section{Introduction}

In the first paper (paper I) \cite{To12} of this series of two
papers, we obtained a number of analytical results concerning the
continuum percolation of overlapping hyperspheres and overlapping
oriented hypercubes that applied across all Euclidean space
dimensions. Among other results, it was shown analytically that
certain lower-order Pad{\'e } approximants on the mean cluster
number $S$ are lower bounds on $S$ for both hyperspheres and
hypercubes in $d$-dimensional Euclidean space $\mathbb{R}^d$ and
that they become exact asymptotically as $d \rightarrow \infty$.
In this asymptotic limit, the dimensionless density at percolation
$\eta_c$ tends to $2^{-d}$. An important consequence of the
aforementioned analysis is that this large-$d$ percolation value
is an important contribution to the low-dimensional percolation
value. In other words, low-dimensional results encode
high-dimensional information. Percus-Yevick-like approximations
for the cluster number $S$ were also obtained that also become
asymptotically exact as $d \rightarrow \infty$. The analysis was
aided by a striking {\it duality} between the equilibrium
hard-hypersphere (hypercube) fluid system and the continuum
percolation models of overlapping hyperspheres (hypercubes),
namely,
\begin{equation}
P(r;\eta)=-h(r;-\eta)
\label{dual}
\end{equation}
where $P(r;\eta)$ is the pair connectedness function at some
radial distance $r$  and reduced density $\eta$ for the continuum
percolation models and $h(r;\eta)$ is the total correlation
function for the equilibrium hard-particle models.\cite{footnote1}
It was shown that the large-$d$ percolation threshold $\eta_c$
of overlapping hyperspheres and hypercubes
is directly related to the large-$d$ freezing-point density
of corresponding equilibrium hard-particle models.\cite{To12}
The extension of these results for overlapping hyperspheres
and hypercubes to  the case of overlapping
particles of general {\it anisotropic} shape in $d$ dimensions
with a specified orientational probability distribution was also
described.

The bounds and approximations reported in paper I were applied to
assess the accuracy of previous computer simulation results for
$\eta_c$ that span dimensions up to $d=20$ in the case of
overlapping hyperspheres \cite{Kr03,Wa06} and up to $d=15$ in the
case of hypercubes.\cite{Wa06} It is convenient to restate the
best lower bound on $\eta_c$ that was found in Ref. \onlinecite{To12},
namely,
\begin{equation}
\eta_c \ge \frac{\displaystyle 1+ \frac{C_3}{
2^{2d}}}{\displaystyle 2^d\left[1+ \frac{2C_3}{
2^{2d}}+\frac{C_4}{ 2^{3d}} \right]}, \label{eta-2-1}
\end{equation}
where $C_3$ and $C_4$ are the trimer and tetramer statistics
defined and computed as a function of $d$ in paper I. Comparison
of this lower bound to  Kr{\" u}ger's simulation data for
hyperspheres revealed that the bound became progressively tighter
as $d$ increased and became an excellent estimate for $d \ge 7$.
Since it becomes increasingly challenging to estimate percolation
thresholds from simulations in high dimensions, it was not
surprising that even Kr{\" u}ger's high-quality simulation data
fell slightly below the  lower bound (\ref{eta-2-1}) for $8 \le d
\le 11$. The analytical results of paper I revealed that the
simulation data reported in Ref. \onlinecite{Wa06} for both
hyperspheres and hypercubes were considerably more problematic. In
particular, the authors reported incorrectly that the quantity
$2^d\eta_c$ for these systems were nonmonotonic in dimension and
that hyperspheres have lower thresholds than hypercubes in higher
dimensions while the reverse is true in lower dimensions.

One of the main purposes of the present paper is to exploit the
accuracy of the lower bound (\ref{eta-2-1}) to provide an
efficient simulation method, called  the {\it rescaled-particle}
algorithm, to estimate continuum percolation thresholds across
many dimensions.  Another objective is to obtain an upper bound on
$\eta_c$ as well as to provide additional analytical results for
certain cluster statistics. In Sec. \ref{anal},  we derive these
analytical results. In Sec. \ref{sim}, we describe in detail the
rescaled-particle method, which is applicable for general
continuum percolation models (e.g., spherical and nonspherical
particle shapes). Results for the thresholds of both overlapping
hyperspheres and oriented hypercubes for $ 3 \le d \le 11$ are
compared to upper and lower bounds on $\eta_c$ in Sec.
\ref{results}. In Sec. \ref{conclusions}, we summarize our
conclusions and discuss future work.

\section{Additional Analytical Results}
\label{anal}

\subsection{Basic Definitions}
\label{sub}

A prototypical continuum percolation model consists of equal-sized
overlapping (Poisson distributed) hyperparticles in $\mathbb{R}^d$
at number density $\rho$; see paper I and numerous
references therein.  It is convenient to introduce
the reduced number density $\eta$, defined by the relation
\begin{equation}
\eta=\rho v_1,
\label{eta}
\end{equation}
where $v_1$ is the $d$-dimensional volume of a  hyperparticle; see
Ref. \onlinecite{To12} for an explicit expression of this quantity
for a hypersphere, for example. A cluster statistic that has been
considered by various investigators in one, two and three
dimensions for overlapping spheres is $n_k$, the average number of
$k$-mers per unit number of particles. \cite{Ha77,Co77,Se88,Qu96}
This $k$-mer statistic obeys the following constraint
\begin{equation}
\sum_{k=1} k n_k=1,
\end{equation}
where it is to be noted that $p_k \equiv k n_k$ is the probability that a
given particle is part of a $k$-mer.\cite{Qu96} In what follows, we will derive estimates for $n_k$ and related cluster statistics
for arbitrary dimension.

It was shown in Ref. \onlinecite{Qu96} that $n_k$ can be
explicitly expressed for any $d$ for overlapping hyperspheres as
certain multidimensional integrals involving exponentials whose
arguments contain the union volume of $k$ ``exclusion spheres.''
Two spheres of radius $D/2$ are considered to be connected if they
overlap, i.e., if the center of one lies within a spherical
``exclusion'' region of radius $D$ centered around the other
sphere (see Fig. 3 of paper I). For example, for $k=1$, $k=2$ and
$k=3$, we have
\begin{eqnarray}
n_1& =& \exp[-2^d \eta] \label{n1} \\
n_2 &=& \frac{\rho}{2} \int_{\mathbb{R}^d} \exp[-\rho v_2(r;D)] f(r) d{\bf r} \label{n2}\\
n_3 &=& \frac{\rho^2}{6} \int_{\mathbb{R}^d} d{\bf r}_{12} \int_{\mathbb{R}^d} d{\bf r}_{13}
     \;f(r_{12}) f(r_{23}) \exp[-\rho v_3(r_{12},r_{13},r_{23}; D)] \nonumber \\
&+&
      \frac{\rho^2}{3} \int_{\mathbb{R}^d} d{\bf r}_{12} \int_{\mathbb{R}^d} d{\bf r}_{13}
      \exp[-\rho v_3(r_{12},r_{13},r_{23}; D)], \label{n3}
\end{eqnarray}
where $v_n({\bf r}_1,{\bf r}_2,\ldots, {\bf r}_n;D)$ is the union volume
of $n$ spheres of radius $D$ centered at positions ${\bf r}_1,{\bf r}_2,\ldots, {\bf r}_n$,
${\bf r}_{ij}= {\bf r}_j - {\bf r}_i$ and $r_{ij}= |{\bf r}_{ij}|$ ($i \neq j$).
Moreover,  the radial function $f(r)$ defines the connectedness criterion, i.e.,
\begin{equation}
f(r) =\Theta(D-r),
\label{mayer-sphere}
\end{equation}
and
\begin{equation}
\Theta(x) =\Bigg\{{1, \quad x \ge 0,\atop{0, \quad x <0}}
\end{equation}
is the Heaviside step function.
Note that the factor $2^d$, appearing in Eq. (\ref{n1}),  is the ratio of the exclusion volume $v_{ex}$ to the
volume of a sphere. By virtue of the fact that the spheres are  Poisson distributed in space, it
follows that the mean number of overlaps per sphere $\cal N$ is given by
\begin{equation}
{\cal N}= \rho v_{ex}= 2^d\eta.
\label{N}
\end{equation}

The {\it average number of clusters per unit volume} $\rho_c$ is
directly related to the sum over $n_k$,  \cite{Qu96} namely
\begin{equation}
\frac{\rho_c}{\rho} =\sum_{k=1}^\infty n_k.
\end{equation}
Note that the ratio $\rho_c/\rho$ is the {\it average number of
clusters per particle} and its derivative with respect to $\rho$
(or $\eta$) determines the contact value of the {\it blocking
function} $B(r)$, \cite{Gi90} which is directly related to the
conditional probability of finding two particles belonging to
different clusters separated by distance r, given that one of the
particles is at the origin. For example, in the case of
hyperspheres of diameter $D$, the three-dimensional expression
given in Ref. \onlinecite{Gi90} generalizes as follows:
\begin{equation}
\frac{ d (\rho_c/\rho)}{d \eta} = - 2^{d-1} B(D).
\label{block1}
\end{equation}
We recall here that for overlapping hyperspheres, the blocking function $B(r)$
can be obtained immediately from the pair connectedness function $P(r)$
for any radial distance $r$ via the relation \cite{Co77}
\begin{equation}
P(r) + B(r) =1 \;.
\end{equation}
The fact that $P(r)$ is bounded in the interval $[0,1]$, implies
the same bounds on $B(r)$.

The {\it average cluster number} $Q$ is the average number of
particles in a randomly chosen cluster and is the inverse of
$\rho_c/\rho$, \cite{Qu96} namely,
\begin{equation}
Q = \frac{\rho}{\rho_c}=\left( {\sum_{k=1}^\infty n_k} \right)^{-1}.
\label{defQ}
\end{equation}
This is to be distinguished from the cluster number $S$ (average number of particles in the
cluster containing a randomly chosen particle), which is related to the second moment
of $n_k$: \cite{Ha77}
\begin{equation}
S = \sum_{k=1}^\infty k^2 n_k, \qquad \eta < \eta_c.
\label{defS}
\end{equation}
Unlike $S$, which can also be expressed in terms of the pair
connectedness function $P(r)$, \cite{To12} the average cluster number
$Q$  does not diverge at the percolation threshold when $d \ge 2$.
\cite{Qu96}

\subsection{Estimates of Cluster Statistics and Upper Bounds on $\eta_c$ and ${\cal N}_c$}
\label{bounds}

It was shown in Ref. \onlinecite{To12} that the mean number of overlaps per sphere at the
threshold ${\cal N}_c$ [cf. (\ref{N})] tends to unity as $d \rightarrow \infty$, i.e.,
\begin{equation}
{\cal N}_c \equiv \eta_c \frac{v_{\mbox{\scriptsize ex}}}{v_1} \sim 1, \qquad d \rightarrow \infty,
\end{equation}
which applies to spherical as well as nonspherical particles (with
specified orientational distribution). This does not mean that the concentrations
of  monomers, dimers, trimers, etc. at the threshold are negligibly small,
even if finite clusters become more ramified as the space dimension grows. \cite{To12}
To explicitly prove this property, we first observe that the exact formula
for $n_1$ [cf. (\ref{n1})] together with the exact asymptotic result
\begin{equation}
\eta_c \sim \frac{1}{2^d}, \qquad d \rightarrow \infty,
\label{eta_c}
\end{equation}
which applies to any oriented centrally symmetric particle (e.g., spheres, cubes, ellipsoids, etc.) \cite{To12},
implies the following asymptotic result
\begin{equation}
n_1 \sim \exp(-1) = 0.3678794\ldots, \qquad d \rightarrow \infty.
\end{equation}

We will now show that as $d$ becomes large, monomers, as opposed to any $k$-mer for $k\ge 2$,
are dominant in so far as concentration is concerned, i.e., $n_1$ is appreciably larger than $n_2$
and therefore is appreciably larger than $n_k$ with $k \ge 3$, since $n_k > n_{k+1}$ for any positive but bounded $\eta$.
Let us begin with the formula (\ref{n2}) for the dimer statistic $n_2$, which we can
rewrite as follows:
\begin{equation}
n_2 = d 2^{d-1} \eta \exp[-2^{d+1} \eta] \int_0^D r^{d-1} \exp[2^d \eta \alpha(r;D)],
\label{n2-2}
\end{equation}
where
\begin{equation}
\alpha(r;R) = \frac{v^{\mbox{int}}_2(r;R)}{v_1(R)}= \frac{2\Gamma(1+d/2)}{\sqrt{\pi})\Gamma((d+1)/2)}\int_0^{\cos^{-1}(r/(2R)}
\sin(\theta)^d d\theta
\end{equation}
Here we have used the fact that $v_2(r;R)= 2 v_1(R)- v^{\mbox{int}}_2(r;R)$, where the latter
quantity is the intersection volume of two spheres of radius $R$ whose
centers are separated by the distance $r$. The dimensionless intersection volume $\alpha(r;R)$,
which has support in the interval $[0,2R]$,
has been explicitly given for any $d$ in a variety of representations \cite{To12,To06} and played
an important role in paper I. Now since $\alpha(D;D)$ decays to zero exponentially fast
according to the asymptotic relation \cite{To06}
\begin{equation}
\alpha(D;D) \sim \left(\frac{6}{\pi}\right)^{1/2}  \left(\frac{3}{4}\right)^{d/2}
\frac{1}{d^{1/2}},
\label{alpha-asym}
\end{equation}
it immediately follows from (\ref{n2-2}) and (\ref{eta_c}) that
\begin{equation}
n_2 \sim \frac{\exp(-2)}{2} = 0.06766764\ldots, \qquad d \rightarrow \infty,
\label{n2-3}
\end{equation}
and hence we find
\begin{equation}
\frac{n_1}{n_2} \sim  2\exp(1)= 5.436563\ldots, \qquad d \rightarrow \infty,
\end{equation}
which is what we set out to prove.

We now derive lower bounds on $n_k$ for $k \ge 2$ as a function of $d$
for any $\eta$. Let us begin with the case $k=2$. Since $\alpha(r;D)$ is a monotonically
decreasing function of $r$, we have that $\alpha(D;D) \le \alpha(r;D)$
in the interval $[0,D]$ and hence combined with the exact formula (\ref{n2-2})
yields the lower bound
\begin{equation}
n_2 \ge 2^{d-1} \eta \exp[-2^{d+1} \eta]  \exp[2^d \eta \alpha(D;D)].
\label{n2-bound}
\end{equation}
It is noteworthy that in light of (\ref{alpha-asym}), the lower bound (\ref{n2-bound}) becomes
asymptotically exact in the high-$d$ limit, i.e., we recover (\ref{n2-3}).
Using similar arguments and the formulas for $n_k$ given in Ref. \onlinecite{Qu96},
we obtain the following generally weaker lower bounds on $n_k$ for any $k$:
\begin{equation}
n_k \ge \frac{2^{(k-1)d}}{k(k-1)!} \eta^{k-1} \exp[-k 2^d \eta].
\end{equation}
While for $k=1$, this bound is exact, it is weaker than (\ref{n2-bound}) for $k=2$.

Through second order in $\eta$, formulas (\ref{n1}), (\ref{n2}) and
(\ref{n3}) for $n_1$, $n_2$ and $n_3$, respectively, yield
\begin{eqnarray}
n_1 &=& 1 -2^d \eta+2^{2d-1} \eta^2 + {\cal O}(\eta^3) \\
n_2 &=& 2^{d-1} \eta - \left(2^{2d} +\frac{C_3}{2}\right) \eta^2 + {\cal O}(\eta^3) \\
n_3 &=&  \left(2^{2d-1} +\frac{C_3}{3}\right) \eta^2 + {\cal O}(\eta^3)
\end{eqnarray}
Similarly, using expression (\ref{defQ}) and the relations immediately above,
we can obtain the corresponding density expansion for $Q$:
\begin{equation}
Q=1+  2^{d-1} \eta +\left(2^{2(d-1)} +\frac{C_3}{6}\right) \eta^2 + {\cal O}(\eta^3)
\end{equation}

It was noted in Ref. \onlinecite{To12} that the pole of the [1,1] Pad{\' e} approximant
of the density expansion of $Q$ for $d=3$ yielded an upper bound on the
threshold $\eta_c$ for overlapping spheres.  For general $d$, the [1,1] Pad{\' e} approximant
for $Q$ for either overlapping hyperspheres or oriented hypercubes is given by
\begin{equation}
Q_{[1,1]} \approx \frac{1 - \frac{C_3}{3 \cdot 2^{d} }\eta}
{1-\left[2^{d-1}+ \frac{C_3}{3 \cdot 2^d}\right]\eta},
\label{1-1}
\end{equation}
where the trimer statistic $C_3$ for both models is given in Ref. \onlinecite{To12}.
We now observe that the pole of (\ref{1-1}) is an upper bound on $\eta_c$
for overlapping hyperspheres and oriented hypercubes for any $ d\ge 3$, i.e.,
\begin{equation}
\eta_c \le \frac{1}{2^{d-1}\left[1+ \frac{C_3}{6 \cdot 2^{2(d-1)}}\right]}, \qquad d \ge 3.
\label{upper}
\end{equation}
Using the same methods described in paper I, it is straightforward
to prove that for sufficiently large $d$ and  any $\eta < \eta_c$, relation (\ref{1-1}) bounds
$Q$ from below and hence (\ref{upper})
is a rigorous upper bound on the threshold. This is consistent with the
observation (noted in Sec. \ref{sub}) that the actual function $Q$ does not diverge when the mean cluster
number $S$ diverges, i.e., when $\eta \rightarrow \eta_c$.  We will show in Sec. \ref{results}
that the expression (\ref{upper})  bounds the simulation data for the threshold from above for both
hyperspheres and hypercubes for $3 \le d \le 11$. Moreover, according to
Ref. \onlinecite{To12}, the second term within the brackets of inequality (\ref{upper})
goes to zero exponentially fast in the limit $d \rightarrow \infty$, and hence this bound
asymptotically becomes
\begin{equation}
\eta_c \le \frac{1}{2^d}, \qquad d \rightarrow \infty,
\end{equation}
which is the exact asymptotic result. \cite{To12} Since the lower
bound (\ref{eta-2-1}) also becomes exact in this high-$d$ limit,
the bounds (\ref{eta-2-1}) and (\ref{upper}) converge to the exact
asymptotic value of $2^{-d}$.

Finally, we note that combining relation (\ref{defQ}) and
the approximant (\ref{1-1}) gives the following approximation of
$\rho_c/\rho$:
\begin{equation}
\frac{\rho_c}{\rho} \approx \frac{1-\left[2^{d-1}+ \frac{C_3}{3
\cdot 2^d}\right]\eta} {1 - \frac{C_3}{3 \cdot 2^{d} }\eta}.
\label{rho_c}
\end{equation}
Substituting (\ref{rho_c}) into (\ref{block1}) yields the
following the approximation for the contact value of the blocking
function
\begin{equation}
B(D) =1 -P(D)\approx \frac{1}{1- \frac{C_3}{3 \cdot 2^{d}}\eta}.
\label{PB}
\end{equation}
In the Appendix, we provide plots of $B(D)$ and P(D) versus $\eta$ for
selected dimensions.

\section{Efficient Algorithm to Compute $\eta_c$ Across All Dimensions}
\label{sim}

\subsection{Particle-Addition Method}

A commonly employed approach to estimate the percolation threshold
$\eta_c$ for continuum percolation in two and three dimensions is
the particle-addition method. \cite{Ri97b, Ba02, growth3} Starting
from a configuration of a small number of particles with random
positions in the simulation domain subject to periodic boundary
conditions, new particles are added to the domain sequentially
with randomly chosen positions in the simulation domain. Each time
a new particle is added, the largest cluster in the system (i.e.,
the one containing the largest number of particles) is identified
using a burning-algorithm. \cite{burning} This process is repeated
until a system-spanning cluster forms.

Although conceptually intuitive and used across dimensions,
\cite{Kr03, Wa06} this method become progressively less
computationally efficient as the dimension increases. \cite{To12}
First, to obtain accurate estimates of the percolation threshold,
the size of the particles should be much smaller than the linear
extent of simulation domain so that adding a single particle leads
to a very small increase in the reduced density $\eta$. Without
{\it a priori} knowledge of the percolation threshold, one needs
to start with sufficiently dilute particle configurations, i.e.,
with very small $\eta$. Therefore, an extremely large number of
particles needs to be added to the simulation domain until the
system percolates. For each particle addition, one needs to
identify the largest cluster and check whether it spans the
system, which makes the method computationally very expensive.
Moreover, the dynamic nature of particle addition makes it
difficult to implement efficient methods to check for local
particle connectivity (overlaps of pairs of particles, e.g., the
cell method). This problem increases in severity as the dimension
increases.


We note that highly efficient algorithms have been developed for investigating
clustering and percolation properties of
overlapping disks in two dimensions. For example, several
frontier-tracking methods have been devised to provide very precise estimates
of the percolation threshold $\eta_c$ of overlapping disks. \cite{Qu99,Qu00,Qu07}
However, such algorithms cannot be applied in higher dimensions because there is no
analogous localized boundary in the gradient percolation
method for high-dimensional systems. \cite{sapoval}
Other efficient variations of the particle addition method
have been implemented, e.g., particles are added only to a single growing cluster,
and the cluster-size distribution rather than the spanning cluster is used to
obtain $\eta_c$; \cite{growth1} and rapid ``union-find'' methods have been developed
to keep track of the connected clusters as particles are added. \cite{newman}
However, these variations of the particle-addition method, although
much more efficient than the original one, becomes progressively more difficult to apply
as the dimension increases.

\subsection{Rescaled-Particle Method}

Tight rigorous lower bounds on the percolation threshold, such as
the ones derived in Ref. 1, enable us to devise a highly efficient
method to estimate $\eta_c$ for overlapping particles with
arbitrary shapes and orientations in $\mathbb{R}^d$. The basic
idea is to generate initial static particle configurations at a
value of $\eta$ that is taken to be the best lower-bound value,
allowing one to finally arrive at the critical value in a
computationally efficient manner, even in high dimensions.
We note that procedures in which particles are rescaled to
obtain $\eta_c$ have been previously proposed. \cite{vicsek} However, to the best
of our knowledge, our method is the first one to combine the particle
rescaling procedure with the tightest lower bound values to
efficiently and accurately obtain $\eta_c$.

Initially, a Poisson distribution of a large number of points
in the simulation domain is generated. Each point in configuration
is then taken to be the centroid of a particle with a specified
or random orientation and a characteristic particle length scale $\ell_0$ (e.g.,
the diameter of a sphere). The initial value of $\ell_0$ is chosen
such that the reduced density $\eta$ of the system equals the tightest
lower bound value. For each particle $i$, a near-neighbor list (NNL)
is obtained that contains the centroids of the particles $j$
whose distance $D_{ij}$ to particle $i$ is smaller than $\gamma \ell_0$
($\gamma>1$). The value of $\gamma$ generally depends on the
continuum-percolation models of interest. A rule of thumb for a choosing good value of $\gamma$
is that the NNL list only contains particles that overlap at
the percolation threshold. In our simulation for hyperspheres and
hypercubes, we have used $\gamma \in [1.05, 1.5]$, depending on the space dimension.
Then the particle sizes are slowly and uniformly increased by increasing $\ell_0$
(i.e., rescaling the particles),
which leads to an increase of the reduced density by a small amount $\delta \eta$.
After each rescaling, the particles in the NNL are checked for
overlap and the largest cluster in the system is identified.
The process is repeated until a system-spanning cluster forms.

Since static particle configurations and predetermined NNL
are used, the complexity for identifying clusters is significantly
reduced. In addition, as the space dimension increases, it was shown that
the lower bounds derived in paper I become increasingly tighter. \cite{To12} Thus, total number of
rescaling before percolation is achieved is much smaller than
the total number of particle additions, which dramatically
improves the efficiency of the algorithm. Furthermore, the increase of the
reduced density $\delta \eta$ can take arbitrarily small values
when $\eta_c$ is approached, rather than a fixed discontinuous
value determined by the particle size via particle-addition methods.
This smooth approach to the critical value
allows a more accurate estimate of the percolation threshold.


\begin{figure}
\begin{center}
$\begin{array}{c@{\hspace{1.5cm}}c}
\includegraphics[height=6.0cm,keepaspectratio]{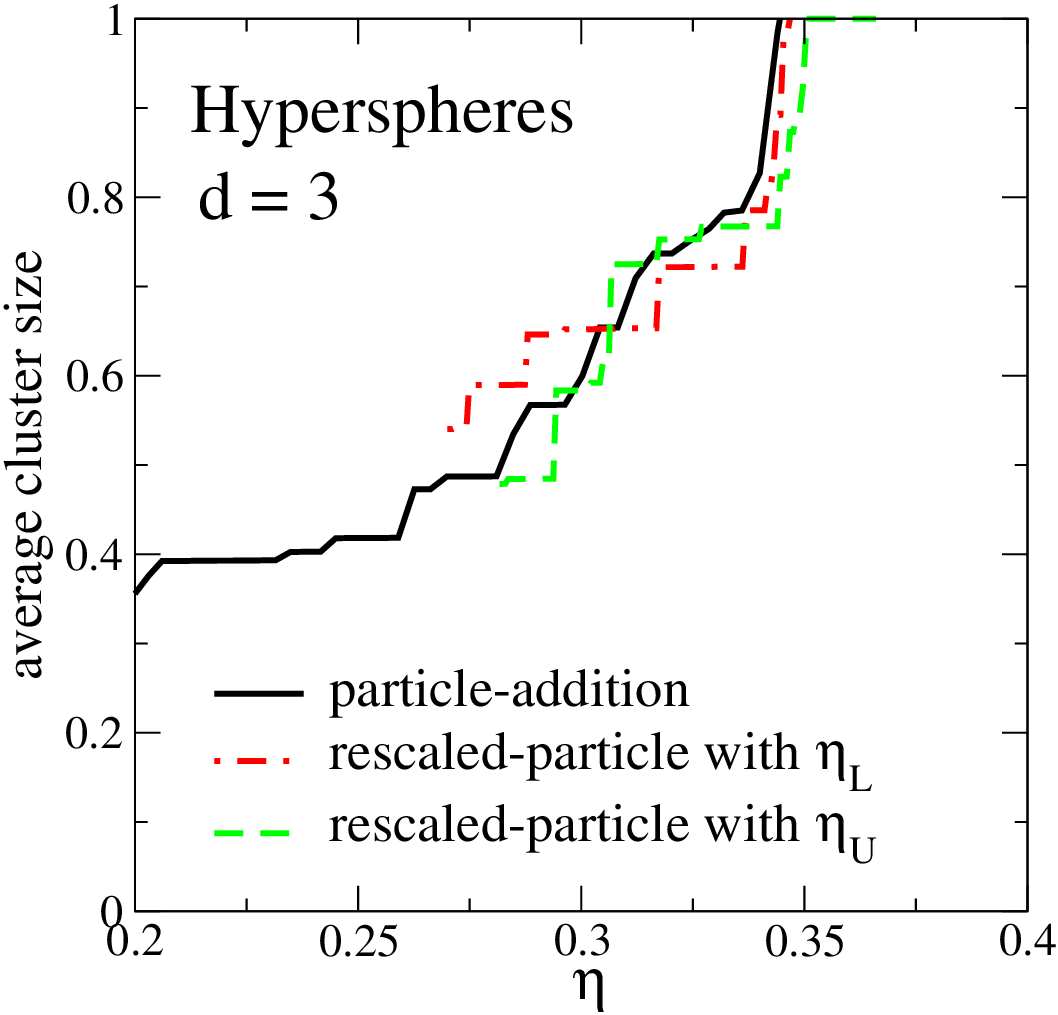} &
\includegraphics[height=6.0cm,keepaspectratio]{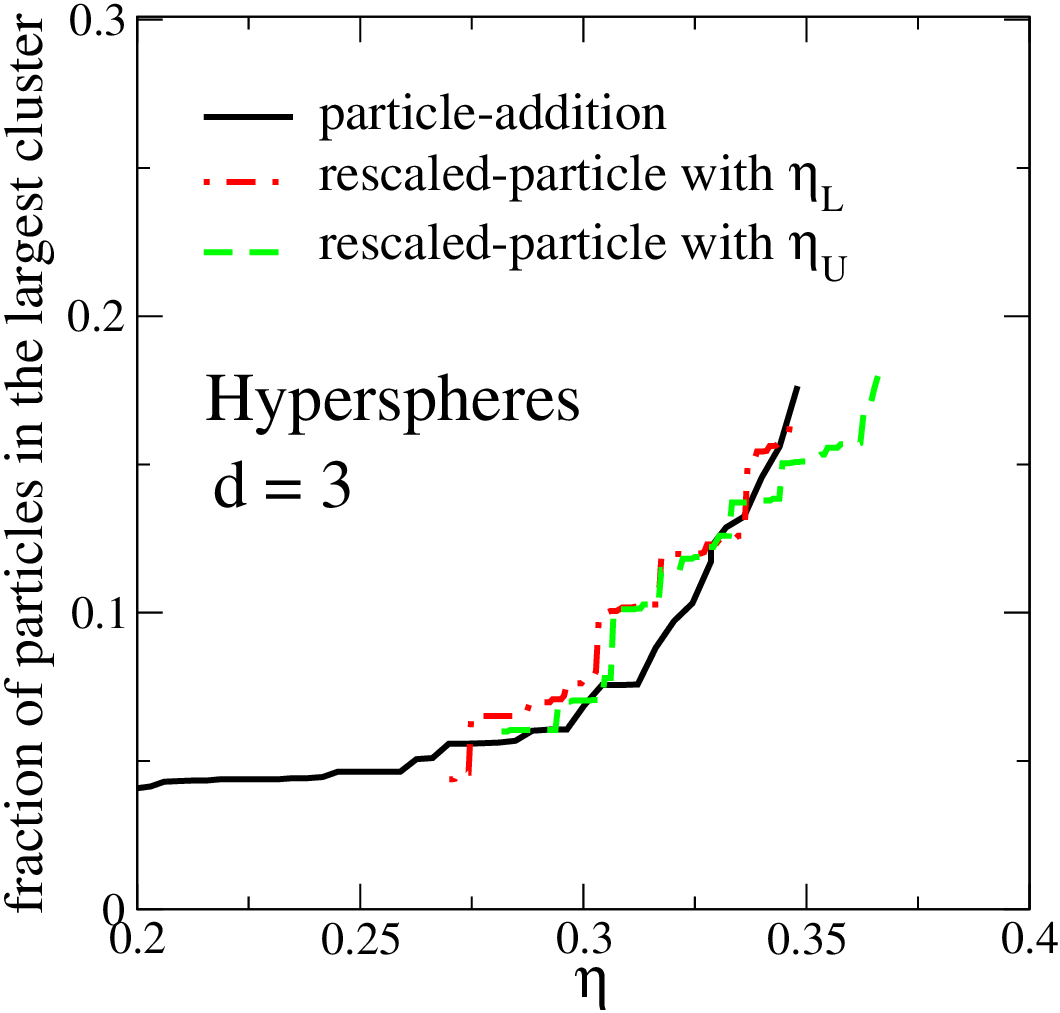} \\
\mbox{(a)} & \mbox{(b)}\\
\end{array}$
\end{center}
\caption{ Cluster statistics, including the linear size of the
largest cluster \cite{fn_cluster} (a) and the fraction of
particles in the largest cluster (b), associated with the
particle-addition and rescaled-particle methods for overlapping
hyperspheres in $\mathbb{R}^3$. The different methods clearly produce
very similar cluster statistics yet the rescaled-particle method
is much more computationally efficient, as explained in the text.}
\label{fig_clusterdynamic}
\end{figure}

Similarly, given an upper bound on $\eta_c$, such as
inequality (\ref{upper}), one can start with a percolated system,
and rescale the particles (i.e., decreasing the particle size
$\ell_0$) to reduce $\eta$. Figure \ref{fig_clusterdynamic} shows
the cluster statistics (e.g., the linear size of the largest
cluster \cite{fn_cluster} and the fraction of particles in the
largest cluster) associated with the particle-addition method and
the rescaled-particle method for overlapping hyperspheres in
$\mathbb{R}^3$. The initial configurations for the
rescaled-particle method include both a non-percolated
configuration with $\eta$ equal to the lower-bound value $\eta_L$
and a percolated configuration with $\eta$ equal to the
upper-bound value $\eta_U$. We note that close to percolation, the
cluster containing the largest number of particles also possesses
the largest linear size.

It is clear that the particle-addition
method and the rescaled-particle method produce very similar
cluster statistics yet the rescaled-particle method is considerably more
computationally efficient in three dimensions. The computational
efficiency improves for overlapping hyperspheres and hypercubes
as the space dimension increases beyond three.
Since the upper bound on $\eta_c$ derived here is not as tight as
the lower bound in relatively low dimensions, we mainly use the rescaled-particle method
starting from non-percolated configurations with $\eta$ equal to
the tightest lower-bound value, as predicted from (\ref{eta-2-1}).
 However, as noted in Sec. \ref{bounds},  as $d
\rightarrow \infty$, both the upper and lower bounds converge to
the exact asymptotic value of $2^{-d}$.



\section{Simulation Results for Overlapping Hyperspheres and Oriented Hypercubes}
\label{results}

\begin{table}[htp]
\centering \caption{Estimates of the percolation threshold
$\eta_c$ for overlapping hyperspheres as obtained from the
rescaled-particle algorithm,  the lower bound (\ref{eta-2-1}), and
the upper bound (\ref{upper}). Also included are the numerical
estimates $\eta_c^*$ of the percolation threshold from a previous study \cite{Kr03}
that satisfy the bounds (\ref{eta-2-1}) and (\ref{upper}). These results
are not reported for $d \ge 8$ because they violate the lower bounds.}
\begin{tabular}{c@{\hspace{1.00cm}}c@{\hspace{1.00cm}}c@{\hspace{1.00cm}}c@{\hspace{1.00cm}}c}\\
\hline\hline
$d$ & $\eta_c^L$ & $\eta_c^*$ & $\eta_c$ & $\eta_c^U$  \\
\hline
2 & 0.748742\ldots & 1.1282 & 1.12810(3) &           \\
3 & 0.271206\ldots & 0.3418 & 0.34289(2) &      0.363636\ldots    \\
4 & 0.111527\ldots & 0.1300  & 0.1304(5)  &      0.167373\ldots     \\
5 & 0.0488542\ldots & 0.0543 & 0.05443(7) &     0.0788179\ldots       \\
6 & 0.0222117\ldots & 0.02346 & 0.02339(5) &     0.0376720\ldots   \\
7 & 0.0103452\ldots & 0.0105 & 0.01051(3) &     0.0181921\ldots  \\
8 & 0.00489917\ldots &   & 0.004904(6) &    0.00885075\ldots  \\
9 & 0.00234800\ldots &  & 0.002353(4) &    0.00432995\ldots  \\
10 & 0.00113534\ldots &  & 0.001138(3) &   0.00212726\ldots  \\
11 & 0.000552682\ldots &  & 0.0005530(3) &  0.00104854\ldots  \\
\hline\hline
\end{tabular}
\label{tab_hypersphere}
\end{table}

\begin{table}[htp]
\centering \caption{Estimates of the percolation threshold
$\eta_c$ for overlapping hypercubes as obtained from the
rescaled-particle algorithm, the lower bound (\ref{eta-2-1}), and
the upper bound (\ref{upper}). Also included are the numerical
estimates $\eta_c^*$ of the percolation threshold from a previous study \cite{Wa06}
that satisfy the bounds (\ref{eta-2-1}) and (\ref{upper}).
These results are not reported for $d \ge 5$ because they violate the lower bounds.}
\begin{tabular}{c@{\hspace{1.00cm}}c@{\hspace{1.00cm}}c@{\hspace{1.00cm}}c@{\hspace{1.00cm}}c}\\
\hline\hline
$d$ & $\eta_c^L$  & $\eta_c^*$ & $\eta_c$ & $\eta_c^U$\\
\hline
2 & 0.732558\ldots & 1.098 & 1.0982(3)  &            \\
3 & 0.256680\ldots & 0.3248 & 0.3247(3) &  0.347824\ldots    \\
4 & 0.103286\ldots & 0.12  & 0.1201(6) &   0.158416\ldots  \\
5 & 0.0447161\ldots &  & 0.05024(7) &   0.0742456\ldots \\
6 & 0.0202386\ldots &  & 0.02104(8) &   0.0354571\ldots  \\
7 & 0.0094301\ldots &  & 0.01004(5) &   0.0171512\ldots  \\
8 & 0.00448213\ldots &  & 0.004498(5) &   0.00837119\ldots  \\
9 & 0.00216025\ldots &  & 0.002166(4) &  0.00411207\ldots  \\
10 & 0.00105159\ldots  &  & 0.001058(4) &  0.00202930\ldots \\
11 & 0.000515602\ldots &  & 0.0005160(3) &   0.00100485\ldots   \\
\hline\hline
\end{tabular}
\label{tab_hypercube}
\end{table}

Using the rescaled-particle method starting from a reduced density
given by the lower-bound estimate (\ref{eta-2-1}), we compute the
percolation threshold for overlapping hyperspheres and oriented
hypercubes in dimensions two through eleven. For each dimension,
different system sizes $N$ are used. Specifically, we employ $N =
10000, ~50000, ~100000$ for $d=2, 3, 4$, $N = 50000, ~100000,
~500000$ for $d=5, 6, 7$, $N = 100000, ~500000, ~1000000$ for
$d=8, 9, 10$, and $N = 1000000, ~2500000, ~5000000$ for $d=11$ and
the results are extrapolated by spline fitting the
finite-system-size data in a log-log plot to obtain the
infinite-system-size estimate of $\eta_c$. For each system size,
the percolation threshold is obtained by averaging over 1000
independent particle configurations for $d=2$ and 3, 500
independent particle configurations for $4 \le d \le 8$, and 100
independent particle configurations for $9 \le d \le 11$.

\begin{figure}[htp]
\begin{center}
$\begin{array}{c}
\\\\
\includegraphics[height=6.0cm,keepaspectratio]{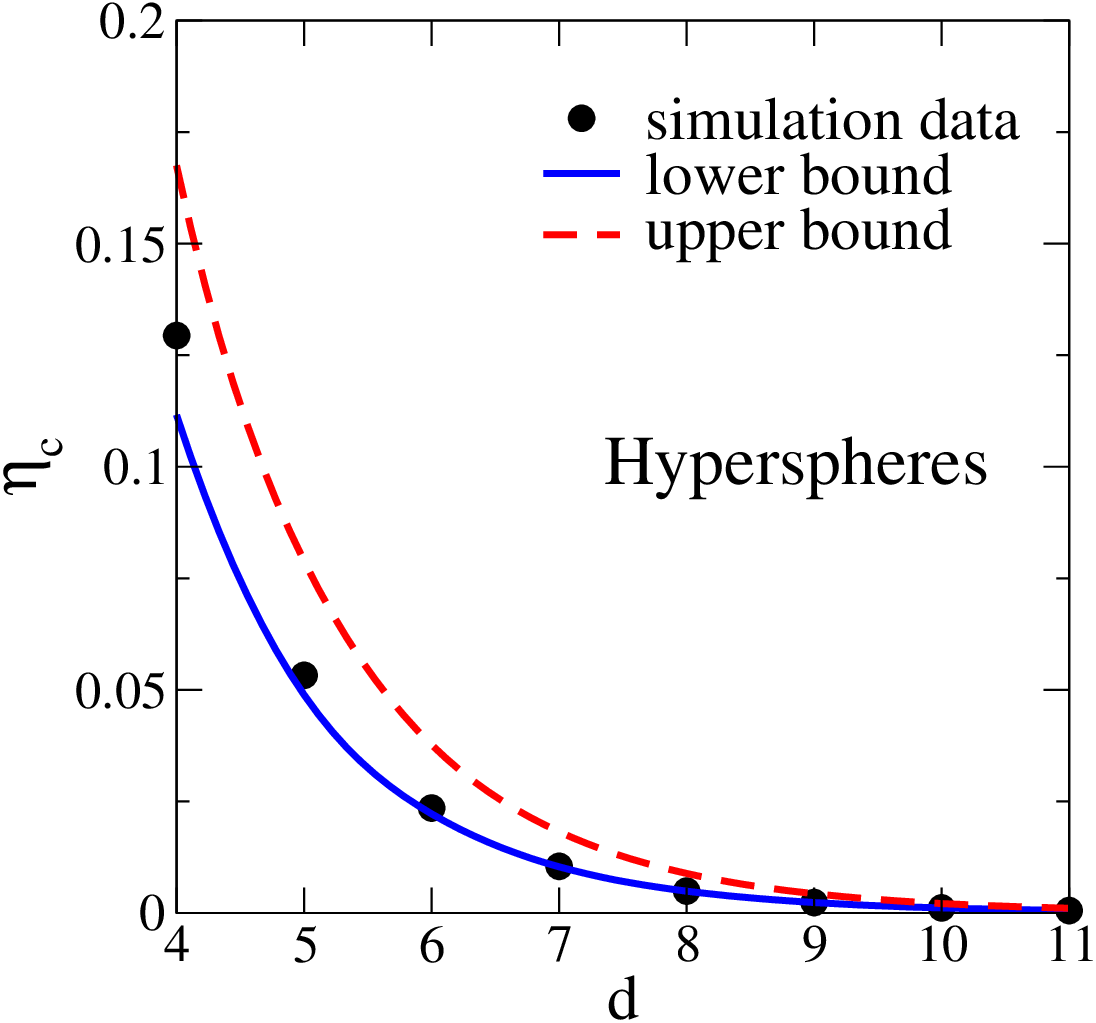}
\end{array}$
\end{center}
\caption{Percolation threshold $\eta_c$ versus dimension $d$ for
overlapping hyperspheres as obtained from the lower bound
(\ref{eta-2-1}), the upper bound (\ref{upper}) and the simulation
data.} \label{fig_bound1}
\end{figure}

\begin{figure}[htp]
\begin{center}
$\begin{array}{c}
\\\\
\includegraphics[height=6.0cm,keepaspectratio]{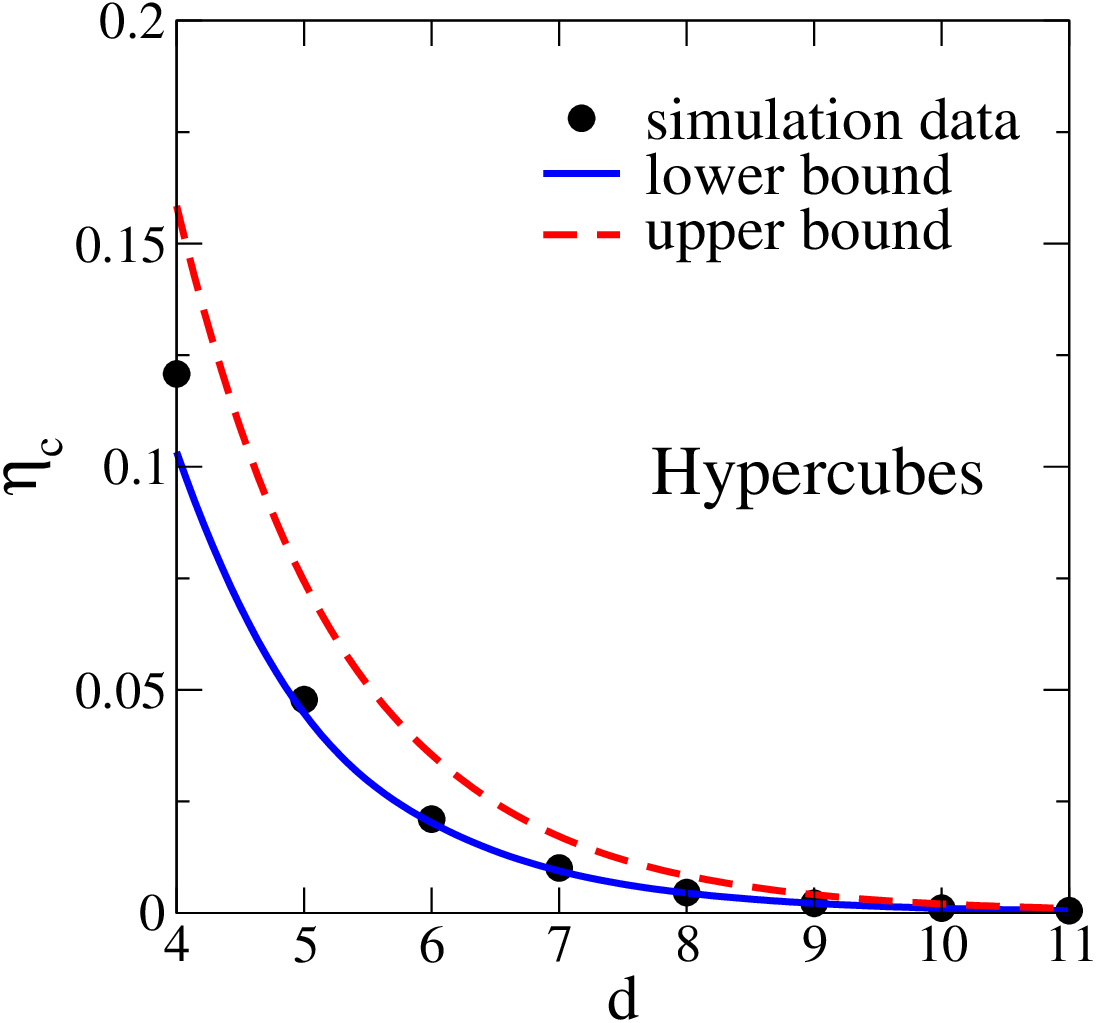}
\end{array}$
\end{center}
\caption{Percolation threshold $\eta_c$ versus dimension $d$ for
overlapping hypercubes as obtained from the lower bound
(\ref{eta-2-1}), the upper bound (\ref{upper}) and the simulation
data.} \label{fig_bound2}
\end{figure}

The obtained percolation threshold values $\eta_c$ for overlapping
hyperspheres and hypercubes in dimensions two through eleven are
respectively given in Table 1 and Table 2, and displayed in Figs.
\ref{fig_bound1} and  \ref{fig_bound2}. We also provide in the
tables and figures the corresponding values of the lower bound
(\ref{eta-2-1}) and upper bound (\ref{upper}) on $\eta_c$ for
purposes of comparison. Note that our simulation data lie very
close to the lower-bound values, and that the lower bounds and
data converge quickly to one another as $d$ increases. Moreover,
we include the numerical estimates of $\eta_c$ from previous
simulation studies for hyperspheres \cite{Kr03} and hypercubes \cite{Wa06}
in case they do not violate the bounds (\ref{eta-2-1}) and (\ref{upper}). It can be clearly seen that
our rescaled-particle method yields much more accurate estimates of
$\eta_c$, especially in high dimensions, since we start with particle
configurations that are already very close to percolation.


Following Ref. \onlinecite{To12}, we use the  threshold  estimate obtained
from the $[2,1]$ Pad{\'e} approximant of $S$  as the basis
to obtain accurate analytical approximations for $\eta_c$
that applies across all dimensions for hyperspheres
and oriented hypercubes. Specifically, we  fit the following function to the
simulation data for $2\le d\le 11$:
\begin{equation}
\eta_c \approx \left ( {1+\frac{b_1}{d^2}+\frac{b_2}{d^4}}\right ) \eta^{(2)}_0,
\label{approx}
\end{equation}
where
\begin{equation}
\eta^{(2)}_0 =\frac{\displaystyle 1+ \frac{C_3}{
2^{2d}}}{\displaystyle 2^d\left[1+ \frac{2C_3}{
2^{2d}}+\frac{C_4}{ 2^{3d}} \right]}
\end{equation}
is the pole associated with the [2, 1] Pad{\' e} approximant for
the mean cluster numver $S$, i.e.,  the tightest lower bound for
$\eta_c$, explicitly given by Eq. (119) of paper I. This lower
bound becomes exact for sufficiently large $d$. Therefore, in
agreement with the conclusions of Ref. \onlinecite{To12}, we see
that the high-dimensional percolation behavior  is an important
contribution to the low-dimensional percolation value. In other
words, low-dimensional results encode high-dimensional
information.

We find that $b_1 = 2.45074$ and $b_2 = -1.65036$ for hyperspheres
with correlation coefficient equal to $0.993194$; and $b_1 =
2.57917$ and $b_2 = -2.29755$ for hypercubes with correlation
coefficient equal to $0.992262$. Note that due to the quality of
the available numerical data reported in paper I, the analogous
analytical approximations based on those data for hyperspheres and
hypercubes only used numerical threshold estimates for $2\le d \le
7$ and $2\le d \le 4$, respectively. Thus, the formula
(\ref{approx}) for hyperspheres and hypercubes supersedes in
accuracy the ones provided in Ref. \onlinecite{To12}.

 As pointed out in Ref. \onlinecite{To12}, the numerical
estimates of $\eta_c$ for hypercubes in $5 \le d \le 15$ by Wagner
{\it et al.} \cite{Wa06} violate the tightest lower bound
(\ref{eta-2-1}). In addition, these simulation data are questionable
in high dimensions since the mean number of overlaps per particle
$\cal N$ [defined in Eq.~(\ref{N})] evaluated at the percolation
threshold, i.e., ${\cal N}_c = 2^d \eta_c$ is incorrectly found to
be a nonmonotonic function of $d$. In particular, these authors
found that ${\cal N}_c$ first decreases as $d$ increases for $2\le
d \le 9$ and then increases as $d$ increases for $10 \le d \le
15$. This led to the incorrect conclusion that hyperspheres have
lower thresholds than hypercubes in higher dimensions while the
reverse is true in lower dimensions.

\begin{figure}
\begin{center}
$\begin{array}{c}
\includegraphics[height=6.0cm,keepaspectratio]{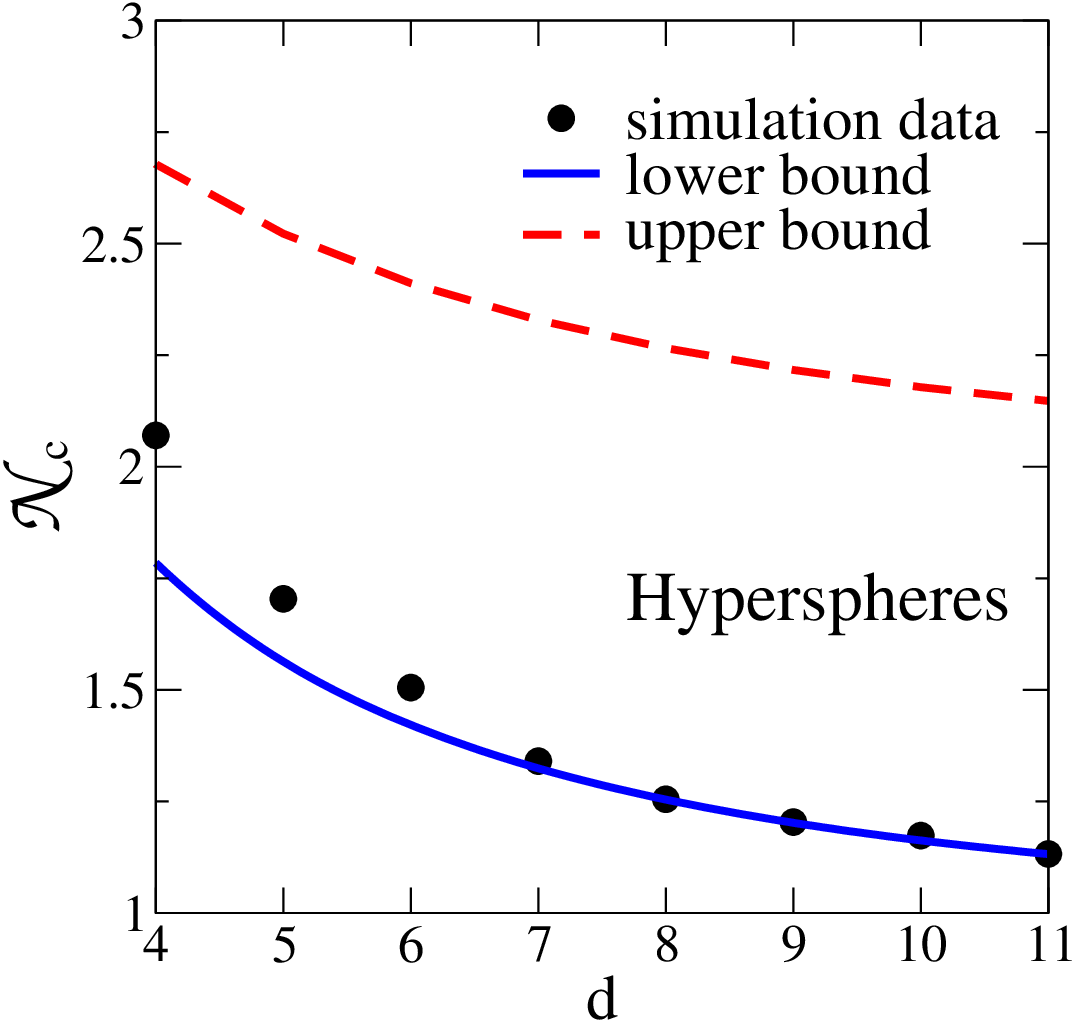}
\end{array}$
\end{center}
\caption{The mean number of overlaps per hypersphere at
percolation threshold ${\cal N}_c = 2^d \eta_c$ as a function of
$d$. Also shown are the quantities $2^d \eta_L$ and $2^d \eta_U$,
where $\eta_L$ and $\eta_U$ are respectively the tightest lower
bound (\ref{eta-2-1}) and upper bound (\ref{upper}). It is clear
that ${\cal N}_c$ is a monotonic function of $d$ and quickly
converges to the asymptotic value of unity as $d$ increases.}
\label{fig_Nc_sphere}
\end{figure}

\begin{figure}
\begin{center}
$\begin{array}{c}
\includegraphics[height=6.0cm,keepaspectratio]{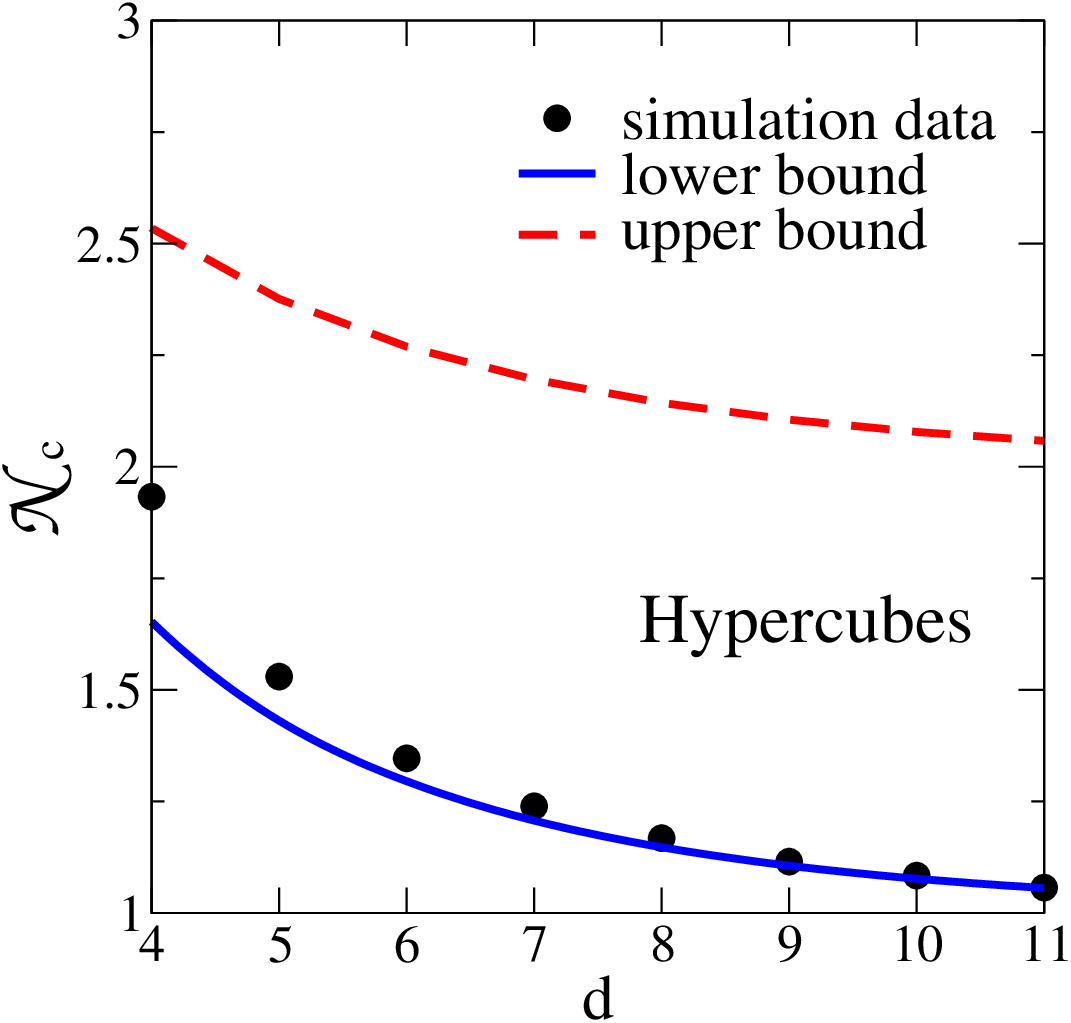}
\end{array}$
\end{center}
\caption{The mean number of overlaps per hypercube at percolation
threshold ${\cal N}_c = 2^d \eta_c$ as a function of $d$. Also
shown are the quantities $2^d \eta_L$ and $2^d \eta_U$, where
$\eta_L$ and $\eta_U$ are respectively the tightest lower bound
(\ref{eta-2-1}) and upper bound (\ref{upper}). It is clear that
${\cal N}_c$ is a monotonic function of $d$ and quickly converges
to the asymptotic value of unity as $d$ increases.}
\label{fig_Nc_cube}
\end{figure}

In Figs. \ref{fig_Nc_sphere} and \ref{fig_Nc_cube}, we show
${\cal N}_c$ as a function of $d$ computed using the estimates of
$\eta_c$ obtained from our simulations for overlapping
hyperspheres and hypercubes, respectively. The quantities $2^d
\eta_L$ and $2^d \eta_U$ are also shown for purposes of
comparison, where $\eta_L$ and $\eta_U$ are respectively the
tightest lower bound (\ref{eta-2-1}) and upper bound
(\ref{upper}). It can be clearly seen that ${\cal N}_c$ for both
overlapping hyperspheres and hypercubes are indeed monotonic
functions of $d$, which quickly converge to the asymptotic value
of unity as $d$ increases. This indicates again that the large-$d$
asymptotic percolation value is an important contribution to the
low-dimensional percolation value. Moreover, one can see from
Tables 1 and 2 that hypercubes always have a lower threshold than
hyperspheres for any fixed finite dimension, and the thresholds of
these two systems approach one another in the limit $d \rightarrow
\infty$, as predicted in Ref. \onlinecite{To12}.

\begin{figure}[htp]
\begin{center}
$\begin{array}{c}
\\\\
\includegraphics[height=6.0cm,keepaspectratio]{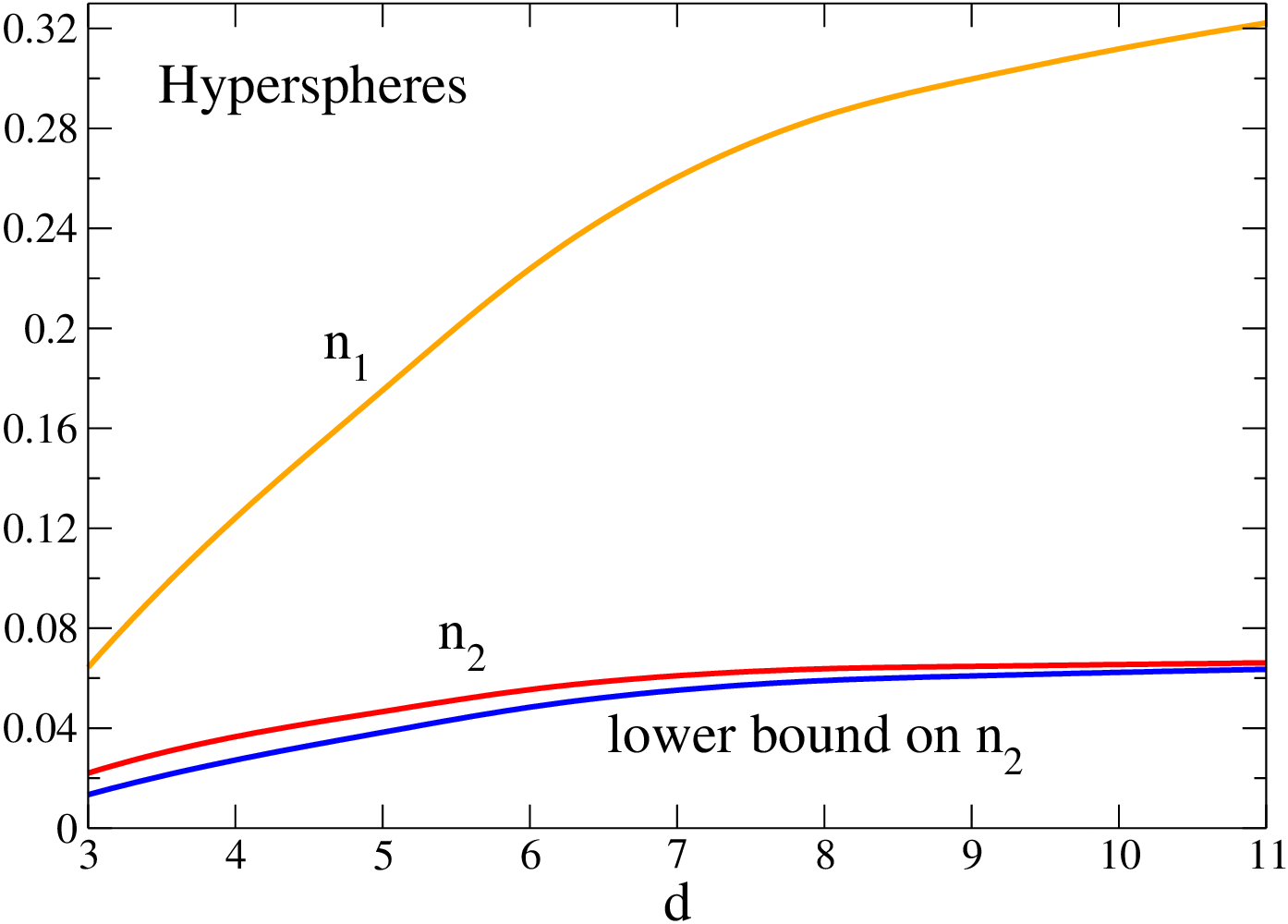}
\end{array}$
\end{center}
\caption{Monomer and dimer concentrations, $n_1$ and $n_2$, at the
percolation threshold $\eta_c$ as a function of  dimension $d$ for
overlapping hyperspheres as obtained from the exact expressions
(\ref{n1}) and  (\ref{n2-2}) and the simulation
data for $\eta_c$ given in Table I. Included in the figure is the lower
bound (\ref{n2-bound}) on $n_2$.} \label{cluster}
\end{figure}

In Fig. \ref{cluster}, we plot the concentration of monomers and dimers, $n_1$ and $n_2$, at the
percolation threshold $\eta_c$ as a function of  dimension $d$ for
overlapping hyperspheres as obtained from the exact expressions
(\ref{n1}) and  (\ref{n2-2}) and the simulation
data for $\eta_c$ given in Table I. Observe that, consistent
with analysis given in Sec. \ref{bounds}, $n_1$ becomes appreciably larger
than $n_2$ (and hence $n_3, n_4$, etc.) as the dimension increases. We include in the figure the lower
bound (\ref{n2-bound}) on $n_2$, which we see becomes tighter
as the space dimension increases, as the analysis of Sec. \ref{bounds} predicts.

\section{Conclusions and Future Work}
\label{conclusions}

We have supplemented the analytical results obtained in paper I by
deriving additional formulas and bounds  for certain cluster
statistics, such as  the concentration of $k$-mers and related
quantities, and obtained an upper bound on the percolation
threshold $\eta_c$. We utilized this upper bound and the tightest
lower bound on $\eta_c$ obtained in paper I  to devise an
efficient simulation method, called  the {\it rescaled-particle}
algorithm, to estimate continuum percolation properties across
many space dimensions.  We applied this simulation procedure here
to compute, with heretofore unattained accuracy,  the threshold
$\eta_c$ and associated mean number of  overlaps per particle
${\cal N}_c$ for both overlapping hyperspheres and oriented
hypercubes for $ 3 \le d \le 11$. Comparison of these simulations
results to corresponding upper and lower bounds on these
percolation properties revealed that the bounds converge to one
another as the space dimension increases. It is noteworthy that
the lower bound provides an excellent estimate of $\eta_c$ and
${\cal N}_c$, even for relatively low dimensions. We confirmed a
prediction of paper I that low-dimensional percolation properties
encode high-dimensional information. We also showed that the
concentration of monomers dominate over concentration values  for
higher-order clusters (dimers, trimers, etc.) as the space
dimension becomes large. Finally, we provided accurate analytical
estimates of the pair connectedness function and blocking function
at their contact values for any $d$ as a function of density.

In paper I, the extension of the continuum percolation results
obtained for overlapping hyperspheres and oriented hypercubes to
cases in which the overlapping hyperparticles are nonspherical
(anisotropic in shape) with some specified orientation
distribution function (e.g., random orientations) was briefly
discussed. Future work will expound on this extension to
overlapping anisotropically-shaped hyperparticles with random
orientations..  The exploration of the generalizations of the
techniques of this series of papers to bound percolation
thresholds in the lattice setting \cite{To12} (bond and site
percolation \cite{burning, SalBook, sahimi1}) represents an
intriguing area for future research.

\section*{Acknowledgements}
\vspace{-0.2in}

This work was supported by the Materials Research Science
and Engineering Center  Program of the National
Science Foundation under Grant No. DMR-0820341 and
 and by the Division of Mathematical
Sciences at the National Science Foundation under Award Number
DMS-1211087.
\bigskip

\noindent{\bf \large Appendix: Approximations for the Blocking and Pair Connectedness Functions  at Contact}
\smallskip

In Sec. II, we noted that the inverse of the cluster number $Q$ is the
of the average number of clusters per unit volume
$\rho_c/\rho$ [see Eq. (\ref{defQ})]. We  derived the approximation (\ref{1-1})
for $Q$, which immediately leads to
the approximation (\ref{rho_c}) for $\rho_c/\rho$ and thus, the
approximation (\ref{PB}) for the contact values of the blocking
function $B(D)$ and pair connectedness function $P(D)$.

\begin{figure}[H]
\begin{center}
$\begin{array}{c}
\\\\
\includegraphics[height=6.0cm,keepaspectratio]{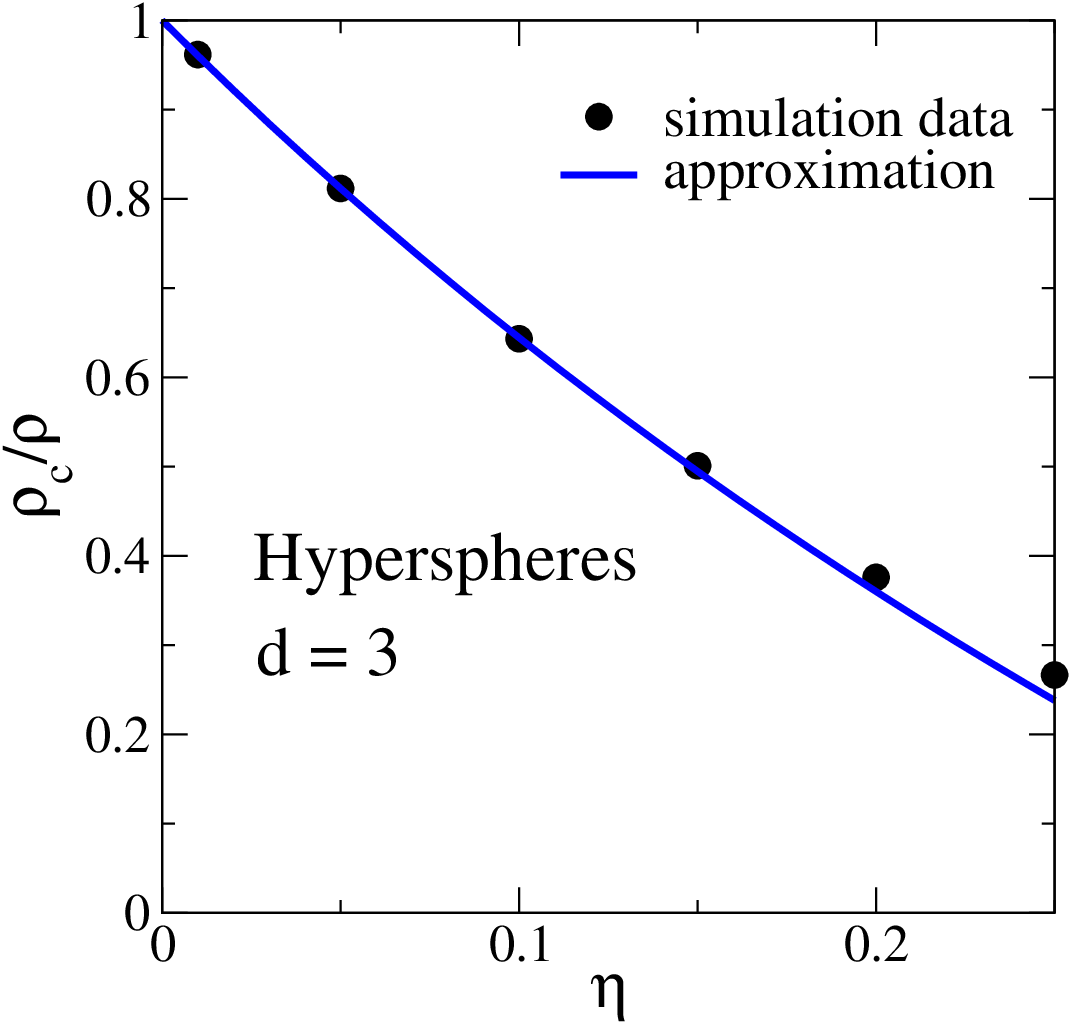}
\end{array}$
\end{center}
\caption{The approximation (\ref{rho_c}) of $\rho_c/\rho$ as a
function of $\eta$ for three-dimensional overlapping spheres. Also
shown are the simulation data reported in Ref. \onlinecite{Gi90}
for purposes of comparison. It can be seen that the approximation
agrees very well with the simulation data  and bounds the data from
below.} \label{fig_rho}
\end{figure}

In this appendix, we compare the approximation (\ref{rho_c}) for $\rho_c/\rho$
versus $\eta$ with
available numerical data for overlapping spheres in
$\mathbb{R}^3$. \cite{Gi90} These results are plotted in Fig. \ref{fig_rho}.
Observe that the approximation agrees with the simulation data very well and bounds the data from
below.

\begin{figure}[H]
\begin{center}
$\begin{array}{c}
\\\\
\includegraphics[height=6.0cm,keepaspectratio]{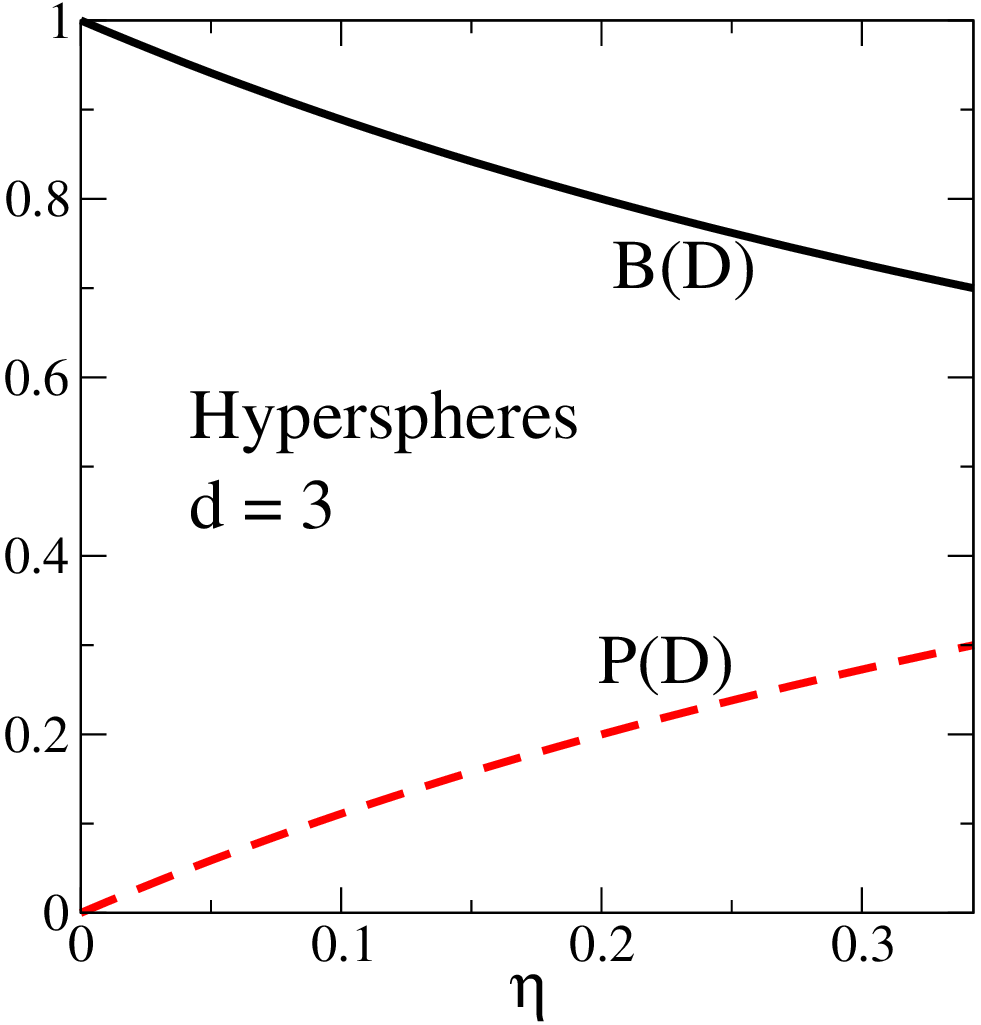}
\end{array}$
\end{center}
\caption{The contact values of the blocking function $B(D)$ and
the pair-connectedness function $P(D)$ versus $\eta$ up to $\eta_c =
0.34289$ (see Table I)
for overlapping spheres in $\mathbb{R}^3$. Note that $B(D)+P(D) =1$
[c.f. (\ref{PB})].}
\label{fig_B_3}
\end{figure}

\begin{figure}[H]
\begin{center}
$\begin{array}{c}
\\\\
\includegraphics[height=6.0cm,keepaspectratio]{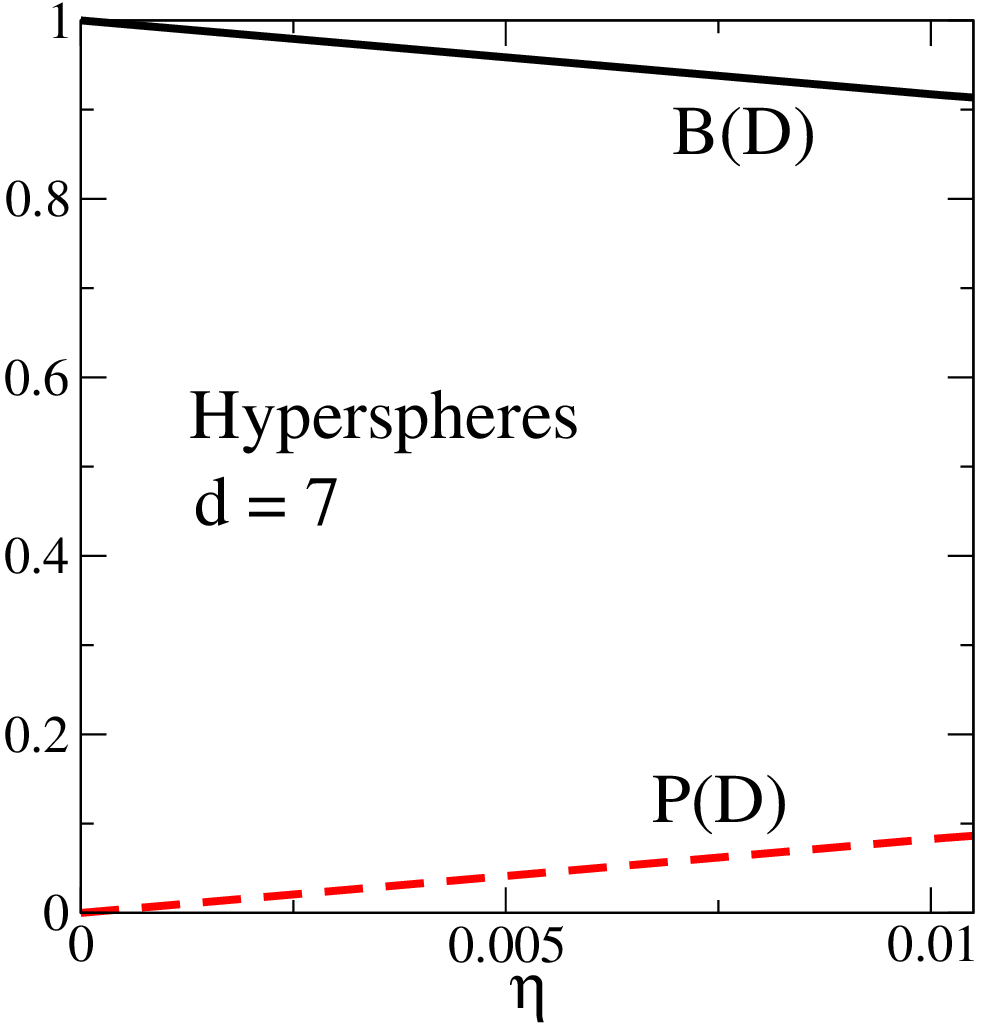}
\end{array}$
\end{center}
\caption{The contact values of the blocking function $B(D)$ and
the pair-connectedness function $P(D)$ versus $\eta$ up to $\eta_c =
0.01051$ (see Table I)
for overlapping hyperspheres in $\mathbb{R}^7$. Note that $B(D)+P(D)
=1$ [c.f. (\ref{PB})].} \label{fig_B_7}
\end{figure}

\begin{figure}[H]
\begin{center}
$\begin{array}{c}
\\\\
\includegraphics[height=6.0cm,keepaspectratio]{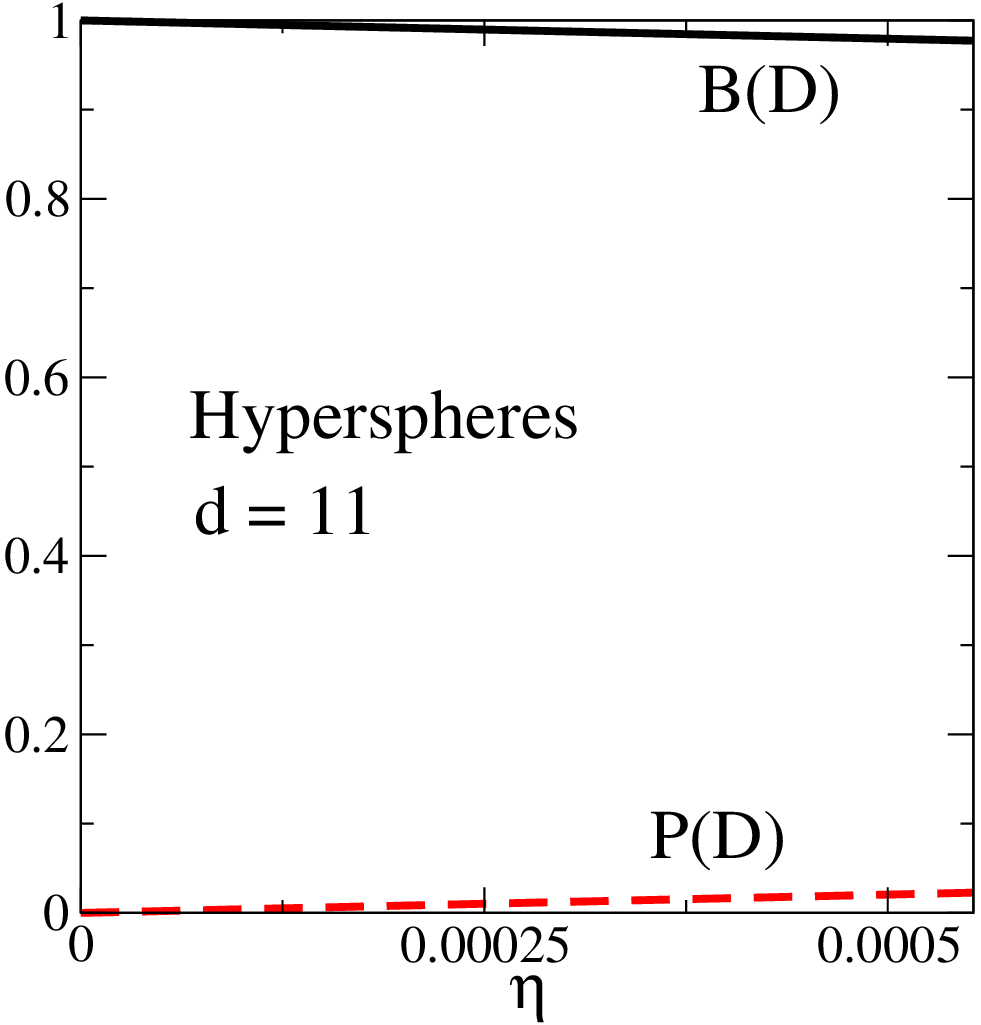}
\end{array}$
\end{center}
\caption{The contact values of the blocking function $B(D)$ and
the pair-connectedness function $P(D)$ versus $\eta$ up to $\eta_c =
0.0005530$ (see Table I)
for overlapping hyperspheres in $\mathbb{R}^{11}$. Note that $B(D)+P(D)
=1$ [c.f. (\ref{PB})].} \label{fig_B_11}
\end{figure}

Since the approximation (\ref{rho_c}) for $\rho_c/\rho$ improves
as the space dimension increases beyond three, we expect that
results derived from it, such as the relation (\ref{PB}) for the
blocking function and pair connectedness function at contact, will
provide accurate approximations across dimensions for $d \ge 4$.
In particular, we provide plots  of the contact values $B(D)$ and
$P(D) = 1-B(D)$ [as predicted by (\ref{PB})] versus $\eta$ for
overlapping hyperspheres in dimensions $3, 7$ and $11$ up to the
respective percolation thresholds in Figs. \ref{fig_B_3},
\ref{fig_B_7} and \ref{fig_B_11}. The functions $B(D)$ and $P(D)$
are equal to unity and zero, respectively, at $\eta=0$ and
decrease and increase monotonically with increasing $\eta$ up to
$\eta_c$. We also see that $B(D)$ and $P(D)$ vary less appreciably
with increasing $\eta$ from the maximum value of unity and minimum
value of zero, respectively, as the space dimension increases.

\begin{thebibliography}{26}
\bibitem{footnote1}
For equilibrium hard-particle systems, the total correlation function $h(r;\eta)$
generally takes on both negative and positive values depending
on the values of the radial distance $r$ and reduced density $\eta$. One can think
of the duality relation (\ref{dual}) as the mapping that is required to convert
a correlation function for a hard-particle system
to a non-negative pair connectedness function (bounded from above by unity) for the corresponding
overlapping particle system.
\bibitem{fn_cluster}
The linear size of a cluster in $\mathbb{R}^d$ is defined to be the largest
distance along any of the $d$ orthogonal directions between the centroids of
any pair of particles in the cluster.
\end{thebibliography}
\end{document}